\documentclass[twocolumn,superscriptaddress,amsmath,aps]{revtex4}

\usepackage{amssymb,mathrsfs,bm,color,times}
\usepackage{graphicx,color}
\usepackage{bm}
\usepackage[bookmarks=false]{hyperref}
\usepackage{amsmath}
\usepackage{subfigure}
\hypersetup{colorlinks=true,citecolor=blue,linkcolor=blue,urlcolor=blue,pdfstartview=FitH,bookmarksopen=true}

\newcommand{\be}{\begin{equation}}
\newcommand{\ee}{\end{equation}}
\newcommand{\bea}{\begin{eqnarray}}
\newcommand{\eea}{\end{eqnarray}}
\newcommand{\bsube}{\begin{subequations}}
\newcommand{\esube}{\end{subequations}}

\newcommand{\la}{\langle}
\newcommand{\ra}{\rangle}

\newcommand{\beq}{\begin{equation}}
\newcommand{\eeq}{\end{equation}}
\newcommand{\beqn}{\begin{eqnarray}}
\newcommand{\eeqn}{\end{eqnarray}}

\newcommand{\bsub}{\begin{subequations}}
\newcommand{\esub}{\end{subequations}}

\begin{document}

\title{Enhanced-sensitivity optical sensing of magnetic fields
using postselected amplification and continuous-wave power recycling}

\author{Yazhi Niu}
\affiliation{Center for Joint Quantum Studies and Department of Physics,
School of Science, \\ Tianjin University, Tianjin 300072, China}

\author{Jialin Li}
\affiliation{Research Center for Quantum Physics and Technologies,
Inner Mongolia University, Hohhot 010021, China}
\affiliation{School of Physical Science and Technology,
Inner Mongolia University, Hohhot 010021, China}

\author{Lupei Qin }
\email{qinlupei@tju.edu.cn}
\affiliation{Center for Joint Quantum Studies and Department of Physics,
School of Science, \\ Tianjin University, Tianjin 300072, China}

\author{Xin-Qi Li}
\email{xinqi.li@imu.edu.cn}
\affiliation{Research Center for Quantum Physics and Technologies,
Inner Mongolia University, Hohhot 010021, China}
\affiliation{School of Physical Science and Technology,
Inner Mongolia University, Hohhot 010021, China}


\date{\today}

\begin{abstract}
Given the importance of precision weak magnetic field measurement, and noting that postselected amplification has not yet been well applied in this area, in this work
we propose to combine the technique of postselected amplification
with continuous-wave power recycling
for enhanced-sensitivity optical sensing of weak magnetic fields.
Based on rigorous derivations, we obtain compact expressions
for the total postselection probability and the signal-to-noise ratio.
We also analyze the effects originating from various non-ideal factors.
In contrast to pulse recycling, continuous-wave recycling has lower control requirements
and is more experimentally feasible.
\end{abstract}

\maketitle

\section{Introduction}

The weak-value amplification (WVA) technique \cite{AAV1} has been widely applied to
various precision measurements, including optical spin Hall effect \cite{Kwi08},
phase shifts \cite{How09a}, beam deflections \cite{How10a}, and frequency shifts \cite{How10b}.
Unfortunately, the amplification effect is achieved by sacrificing photon detection efficiency.
Therefore, the WVA measurement cannot attain higher signal-to-noise ratio (SNR)
than conventional measurement (CM) schemes in general.
Nevertheless, it has been demonstrated that this technique can outperform
conventional measurements under practical constraints,
such as in the presence of beam jitter noise \cite{How14,How15}
and detector saturation \cite{Lun17,ZLJ20}.

To improve the SNR, the idea of power recycling was borrowed for use
in the WVA measurement \cite{Kwiat13,Kwiat15,Kwiat21,Guo16},
while power recycling has been already adopted
in gravitational-wave detection \cite{Dre83,Meer88,Meer91,Dan97,Abb09,Aba11}.
In the interferometric WVA measurement \cite{Kwiat13,Kwiat15,Kwiat21,Guo16},
power recycling collects and reinjects discarded photons into the interferometer.
In this way, in principle, nearly all the incident photons can reach the detector
while preserving the amplified weak value,
thus enhancing the SNR beyond the standard WVA and conventional measurements.

In the context of precision measurement of weak magnetic fields
\cite{Bud07,Bud17,PXH21a,PXH21b,Ger20,Fot20,Bow22,Kaz20},
we proposed a postselection-based scheme
by considering to embed an optical atomic magnetometer
into a Mach-Zehnder interferometer (MZI) \cite{Niu24}.
We also analyzed a pulsed recycling scheme,
which enables the SNR to exceed that of conventional measurement \cite{Niu25}.

\begin{figure}
\centering
\includegraphics[scale=0.9]{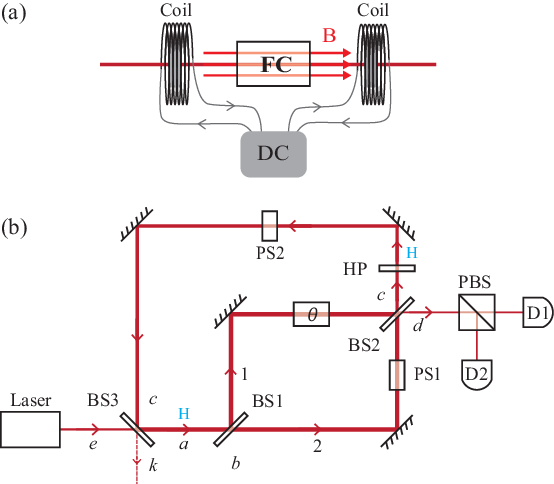}
  \caption{(a) Schematic of an optical magnetometer.
A Faraday crystal (FC) placed between two electric coils
would cause a Faraday rotation (FR) of linearly polarized light,
with the FR angle proportional to the magnetic field to be measured.
(b)
Schematic setup of postselected metrology with continuous-wave power recycling.
The path-state information of the MZI provides the possibility of postselection
via modulating the phase shift through PS1.
The postselected light exits from the port ``d", having an amplified FR angle.
An external loop is designed to recycle the photons exiting from port ``c"
to the input port for re-injection into the interferometer.
Optical elements used in this setup:
beam-splitters (BS1, BS2 and BS3);
polarizing beam-splitter (PBS);
photodetectors (D1 and D2);
phase shifters (PS1 and PS2); horizontal polarizer (HP).  }
\label{fig1}
\end{figure}

In this work, we consider to apply the postselected amplification
and continuous-wave (CW) power recycling techniques
to optical sensing of magnetic fields.
Rather than using WVA, we would like to
use the term postselected amplification,
following Refs.\cite{Llo20,Shu22,Yang23}, from a potentially broader perspective.
The basic idea is also to insert the magnetometer,
which is a Faraday crystal (FC) assumed in this work,
into the MZI as schematically shown in Fig.\ref{fig1}.
Then, the interferometric setup allows not only a path-information selection
for amplification of signal, i.e, the Faraday rotation (FR) angle,
but also coherent multiple interactions of the recycled light with the Faraday crystal.
In steady state, the coherent multiple interactions correspond to
formation of an effective resonant cavity,
between the bright exit port and the laser input port.
As a result, each postselected photon holds an amplified FR angle,
being the same of the single-pass postselected measurement without recycling;
but the total postselected output light intensity
is much stronger than the single-pass postselected light.
Combination of these two advantages can result in an enhanced SNR.
The goal of the present work is to complete the exploration of combining
the FR-based optical magnetometer with laser interferometer
by involving postselection and continuous-wave power recycling.
Compared to pulse recycling, the continuous-wave recycling
requires less demanding light sources and optical elements,
being thus more experimentally accessible.

\section{Setup and single-pass amplification}

We propose to embed the optical magnetometer
into a MZI \cite{Niu24}, as schematically shown in Fig.\ref{fig1}.
The optical magnetometer considered in this work
is a Faraday crystal placed between two electric coils, see Fig.\ref{fig1}(a).
When the left and right circularly polarized lights transmit through the Faraday crystal,
a relative phase between them is introduced.
This is equivalent to a FR of linearly polarized light,
while the FR angle is proportional to the magnetic field to be measured.
The proposed scheme of postselected amplification
together with CW power recycling is displayed in Fig.\ref{fig1}(b).
The path-state information of the MZI provides the possibility of postselection
via modulating the phase shift $e^{i\beta}$ (through PS1).
The postselected light exits from port ``d", holding an amplified FR angle in polarization.
The amplified FR angle can be more reliably measured by two photodetectors (D1 and D2),
in the presence of polarization cross talk of the polarizing beam splitter (PBS) \cite{Niu24}.
An external loop is designed to
recycle the output light existing from the bright port ``c".
In the recycling loop, the horizontal polarizer (HP)
refreshes the polarization of the recycled light,
and the phase shifter PS2 introduces phase compensation,
in order to realize (near) resonance condition.
In front of the input port of the MZI, a beam splitter (BS3) is introduced.
Thus, an effective resonant cavity can be formed between BS2 and BS3,
with the MZI in the forward-propagation path,
and the external loop in the backward-propagation path.

In the absence of recycling (single-pass measurement),
via modulating the phase factor $e^{i\beta}$ (through PS1 in the MZI),
the postselected light exiting from port ``d"
has an amplified FR angle $\tilde{\theta}$, defined by \cite{Niu24}
\bea\label{eq1}
P_{V}=\sin^{2}\theta/(4p_{d1}) = \sin^{2}{\tilde{\theta}} \,.
\eea
$P_{V}$ is the weight of the $|V\ra$ component in the postselected light
(assuming that the input light is in the pure $|H\ra$ state of polarization).
The FR angle $\theta$ is given by $\theta=VBl$, with $B$ the magnetic field to be measured,
$l$ the length of the Faraday crystal, and $V$ the Verdet constant.
The single-pass postselection probability is obtained as
$p_{d1}=(1-\cos\theta\cos\beta)/2$, which depends on the modulation of $\beta$.
From this result, we find that for small $\theta$ and when $e^{i\beta}\to 1$,
the tiny postselection probability would lead to an obvious amplification
of the FR angle, from $\theta$ to $\tilde{\theta}$.
Nevertheless, small $p_{d1}$ means also that the number of the detected photons
$\widetilde{N}_{det}=p_{d1}N$ is largely reduced compared to $N$,
the number of incident photons per unit time.
Thus stronger shot noise would increase the estimation uncertainty of $\tilde{\theta}$,
which is given by $\delta \tilde{\theta}=(2\sqrt{\widetilde{N}_{det}})^{-1}$.
For small $\theta$, we approximately have
$\tilde{\theta}\simeq \theta/(2\sqrt{p_{d1}})$.
We then obtain the SNR as
${\cal R}=\tilde{\theta}/\delta \tilde{\theta}=\theta\sqrt{N}$,
which is the same of the conventional measurement.

\section{Power recycling and main results}

For the convenience of theoretical treatment,
the following evolution and measurement operators can be introduced.
For the FR caused by the Faraday crystal in path ``1"
(with the path state denoted by $|1\ra$),
$\hat{U}_{1}= e^{i\theta |1\ra\la 1| \hat{\sigma}_{x}}$,
with $\hat{\sigma}_{x}= |V\ra\la H|+{\rm h.c.}$,
is introduced to act on the state $|H\ra|1\ra$;
while $\hat{U}_{2}= e^{i\beta|2\ra\la 2|}$
is introduced to act on $|H\ra|2\ra$ in path ``2".
The action of BS3 is described by the scattering matrix (operator)
$\hat{S}_{3}= (t|k\ra\la c|+r|a\ra\la c|)+(t|a\ra\la e|-r|k\ra\la e|)$,
with $t$ and $r$ the transmission and reflection coefficients, which satisfy $t^{2}+r^{2}=1$.
BS1 and BS2, with 50/50 transmission and reflection,
can be similarly described by $\hat{S}_1$ and $\hat{S}_2$
by assigning $t_j=r_j=1/\sqrt{2}$ ($j=1, 2$).
Moreover, a couple of measurement operators can be introduced,
such as $\hat{M}_{H}=|H\ra\la H|$
for polarization filtering through the HP in the recycling loop,
and $\hat{M}_{\nu}=|\nu\ra\la \nu|$ for port selections
(with $\nu=a, c, d$ and $k$ denoting the port information).

With the help of these operators for state propagation,
along the line of derivation presented in detail in Ref.\ \cite{Niu25},
one can obtain the following main results.
In steady state, the state of the total reflecting light in port ``k" is obtained as
\bea\label{eq2}
|\widetilde{\psi}_{r}\ra
= \left[ -r + t(1-r\mathcal{Z})^{-1}(\mathcal{Z} t)\right] |H\ra \,.
\eea
The amplitude of this unnormalized state
corresponds to the total reflection coefficient in port ``k".
The physical meaning of this result is clear by noting that
the multiple-round-trip summation of infinite series
is characterized by the cyclic iteration factor $r\mathcal{Z}$.
Denoting $\mathcal{Z}=A e^{i\varphi}$ and $\varphi = \varphi_{0}+\chi$
(with $\varphi_{0}$ the phase compensation in the recycling loop),
the two key quantities read as
\begin{align}\label{eq3}
\chi &= \arctan\left(\frac{\sin\beta}{\cos\theta+\cos\beta}\right), \nonumber\\
A &= \frac{1}{2}
\sqrt{(1-\gamma)\left(1+\cos^{2}\theta+2\cos\theta\cos\beta\right)}.
\end{align}
In this result, photon loss has been accounted for by the ratio factor $\gamma$.
Similarly, the postselected outgoing state in port ``d" is obtained as
\bea\label{eq4}
|\widetilde{\psi}_{t}\ra
= t(1-r\mathcal{Z})^{-1}\hat{M}_{-} |H\ra \,,
\eea
while the postselection operator is given by
$\hat{M}_{-}=(e^{i\theta\hat{\sigma}_{x}}-e^{i\beta})/2$.
Moreover, the normalization factor of this state gives
the total postselection probability in port ``d", as
\bea\label{eq5}
P_{d}
=\la\widetilde{\psi}_{t}|\widetilde{\psi}_{t} \ra
= t^2 |1-r\mathcal{Z}|^{-2} p_{d1}
\eea
with $p_{d1}$ the single-pass postselection probability,
say, $p_{d1}=(1-\cos\theta\cos\beta)/2$.

Based on Eq.\ (\ref{eq4}), we notice that the signal amplification is the same as that of
the single-pass postselected measurement.
This is clear by noting that the polarization of the postselected light in port ``d"
is determined by $\hat{M}_{-} |H\ra$,
which holds an amplified FR angle $\tilde{\theta}$ given by Eq.\ (\ref{eq1}).
This result is not surprising, since each recycled photon from port ``c" has bee refreshed
to the polarization of $|H\ra$ by the HP in the recycling loop.

Based on Eq.\ (\ref{eq2}), if $\varphi=2\pi n$ (with $n$ an arbitrary integer) and $r=A$,
which is known as {\it impedance matching} condition,
we find that the total reflection amplitude from BS3
in port ``k" vanishes, as a result of
completely destructive interference of all the recycling lights.

On the other hand, based on Eq.\ (\ref{eq5}), under the impedance matching condition,
we find $P_d=p_{d1}/t^2$. Here we have used $t^2=1-r^2$.
Of interest is that this result, under the particular condition,
precisely matches that of the pulse-recycling scheme \cite{Niu25},
say, $P_d=(1-|{\cal Z}|^2)^{-1} p_{d1}$, by noting that $|{\cal Z}|^2=A^2=r^2$.
The simplest understanding of this coincidence is:
Impedance matching in the CW-recycling scheme eliminates input light reflection,
thus creating a scenario analogous to that of the pulse-recycling scheme,
where the polarizing beam splitter (PBS) directs all the recycled light into the input port of the MZI.
However, the two recycling schemes differ drastically in physics.
In the CW-recycling setup,
partial waves in the postselection port superpose coherently.
By contrast, partial waves in the pulse-recycling scheme yield
an incoherent statistical mixture in the same port.
One can check that the individual probabilities of the partial waves
in the coherent superposition are {\it not} equal to
those of the partial waves undergoing incoherent statistical mixing.
Nevertheless, the overall probability resulting from the coherent superposition
matches the total probability of the incoherent mixture.
Quantum interference is the key to yielding this identical total probability.

The above result of identical $P_d$ is merely a special case under impedance matching.
If $r\neq A$ and/or $\varphi\neq 2\pi n$, the total postselection probability $P_d$
differs between the CW-recycling and pulse-recycling setups.

As a consequence of recycling,
the estimate uncertainty of the amplified FR angle $\tilde{\theta}$
is improved as $\delta \tilde{\theta}=(2\sqrt{\widetilde{N}_{det}})^{-1}$,
now with $\widetilde{N}_{det}=P_d N$,
which is much larger than
the single-pass postselected photon number $\widetilde{N}_{det}=p_{d1}N$.
Thus, the rescaled SNR, say, scaled by $\theta\sqrt{N}$ (the SNR without recycling),
is simply given by
\bea\label{eq6}
\widetilde{R}_{S/N} = \sqrt{P_d/p_{d1}} \,.
\eea
Therefore, the key quantity to be analyzed is the total postselection probabilty $P_d$.
Below we show numerical results for $P_d$ and $\widetilde{R}_{S/N}$,
by accounting for impedance mismatch and photon loss in the recycling loop,
and in particular, making comparison with those from the pulse-recycling scheme \cite{Niu25}.

\begin{figure}
\centering
\includegraphics[scale=0.85]{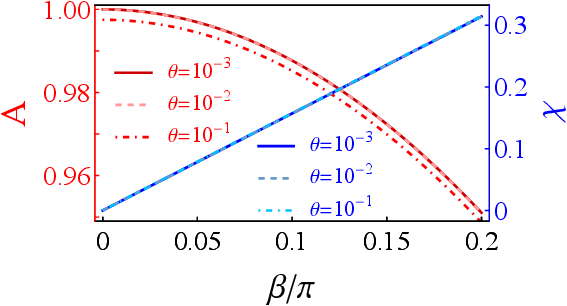}
\caption{
Variations of the effective reflection amplitude $A$ and phase $\chi$
{\it versus} the postselection parameter $\beta$. }
\label{fig2}
\end{figure}

\section{Numerical demonstration of non-ideal effects}

To account for the effect of impedance mismatch, we show first in Fig.\ref{fig2} the dependence of
the amplitude $A$ and phase $\chi$ of the effective reflection coefficient ${\cal Z}$
on the postselection parameter $\beta$.
Our interest is to estimate weak magnetic fields, which correspond to small values of $\theta$.
For small $\theta$, we approximately have
$A\simeq \sqrt{1-\gamma} \cos(\beta/2)$ and $\chi\simeq \beta/2$.
This can facilitate the setting of
the phase compensation $\varphi_0$ in the recycling loop,
and the design of the reflection coefficient $r$ of BS3.

\begin{figure}
\centering
\includegraphics[scale=0.9]{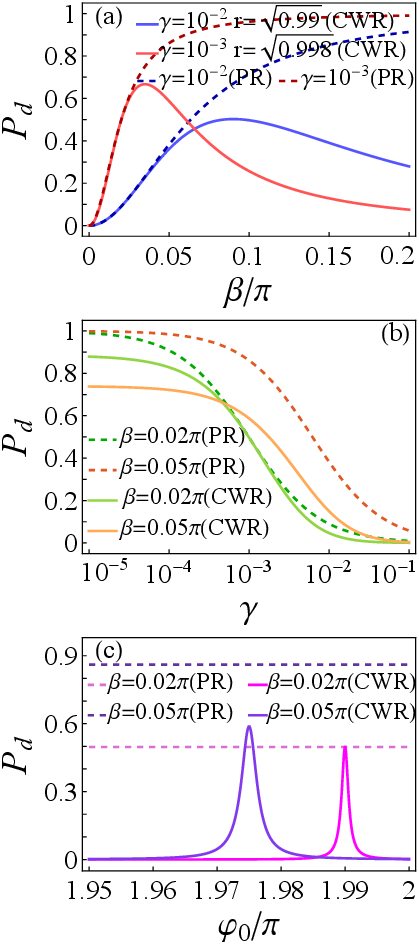}
\caption{
Dependence of the total postselection probability $P_d$
on (a) the postselection parameter $\beta$,
(b) the photon loss $\gamma$, and (c) the compensated phase $\varphi_0$.
In (a)-(c), $\theta=10^{-3}$ is assumed.
Photon loss $\gamma=10^{-3}$ is assumed in (c).
Results for the continuous-wave recycling (CWR)
and pulsed recycling (PR) schemes are compared.    }
\label{fig3}
\end{figure}

\begin{figure}
\centering
\includegraphics[scale=0.9]{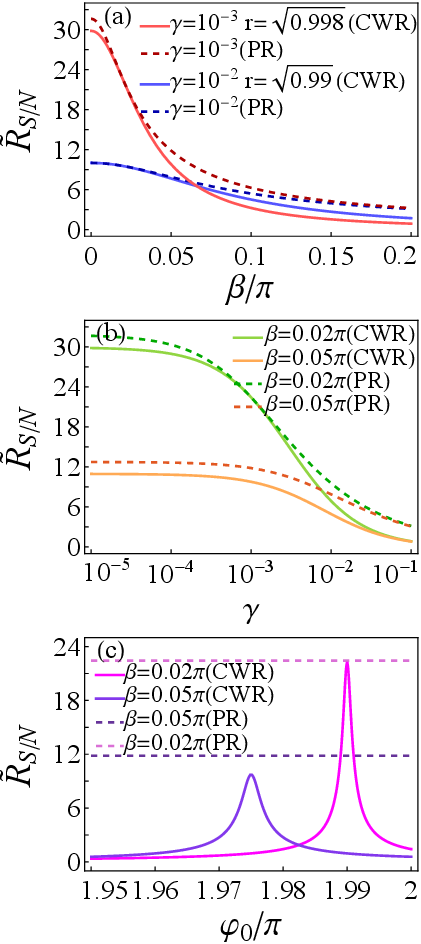}
\caption{
Dependence of the rescaled SNR $\widetilde{R}_{S/N}$
on (a) the postselection parameter $\beta$,
(b) the photon loss $\gamma$, and (c) the compensated phase $\varphi_0$.
In (a)-(c), $\theta=10^{-3}$ is assumed.
Photon loss $\gamma=10^{-3}$ is assumed in (c).
Results for the continuous-wave recycling (CWR)
and pulsed recycling (PR) schemes are compared. }
\label{fig4}
\end{figure}

In Fig.\ref{fig3}, for the MZI-based recycling setup,
we illustrate the dependence of $P_{d}$ on several setup parameters.
In Fig.\ref{fig3}(a), owing to photon loss, we show that $P_{d}$ will decrease
when performing stronger postselection (by setting smaller $\beta$).
In the small $\beta$ regime, the dependence of $P_{d}$ on $\beta$
exhibits identical trends under the CW-recycling and pulse-recycling schemes.
Nevertheless, owing to the amplification of signal,
the overall SNR can be enhanced with tighter postselection (see the results in Fig.\ref{fig4}).
In the larger $\beta$ regime,
$P_{d}$ increases with $\beta$ under the pulse-recycling scheme,
whereas a turnover emerges for the CW-recycling scheme.
The reason for the latter behavior is that a greater impedance mismatch
arises as $\beta$ increases, which leads to more severe photon loss
at the input port (i.e., non-vanishing reflection at BS3).
In Fig.\ref{fig3}(a), we assume $\gamma=10^{-3}$ and $10^{-2}$,
while in Fig.\ref{fig3}(b) we display the continuous behavior with the change of $\gamma$.
Moreover, in Fig.\ref{fig3}(c), we display the behavior
owing to impedance mismatch from phase modulation error,
say, with $\varphi = \varphi_0+\chi$ deviating from $2\pi m$.

In Fig.\ref{fig4}(a),
we show that the SNR can be enhanced by reducing the postselection phase parameter $\beta$,
despite that for small $\beta$ the postselection probability $P_d$ is small
in the presence of photon loss.
This arises from the tradeoff between $P_d$
and signal amplification from $\theta$ to $\tilde{\theta}$.
Generally speaking, the SNR of the pulse-recycling scheme is superior to that of the CW-recycling scheme; however, under full impedance matching conditions, both schemes yield identical SNR,
for instance, as evidenced by the coincidence point of the two curves at $\gamma=10^{-3}$.
Actually, this latter feature has been noticed
based on the analytical analysis in the former part.
In Fig.\ref{fig4}(b), we show degradation of the SNR owing to photon loss.
Noting that the loss ratio $\gamma$ jointly determines
the effective reflection coefficient $A$,
rendering $A$ dependent on $\gamma$,
we thus observe also a coincidence point at $A=r$.
In Fig.\ref{fig4}(c), again,
we illustrate the importance of phase matching condition,
by showing that the SNR peaks at $\phi=\varphi_0+\chi=2\pi m$,
with a narrow window of modulating the compensation phase $\varphi_0$.

\section{Discussion and conclusion}

The signal amplification from $\theta$ to $\tilde{\theta}$, based on Eq.\ (\ref{eq1}),
is determined by the single-pass postselection probability $p_{d1}$,
which can make $\tilde{\theta}/\theta$ large in principle.
Of more practical relevance is the SNR enhancement,
which is governed by the total postselection probability $P_{d}$,
in particular, based on Eq.\ (\ref{eq6}), by the ratio $P_{d}/p_{d1}$.
Thus, the SNR enhancement can be also large in principle.
However, in practice, both the signal amplification and SNR enhancement are subject to limitations
stemming from the fidelity of optical components, and especially photon loss.

We noticed two proof-of-principle experiments, which demonstrated the CW-recycling
and the pulse-recycling schemes \cite{Guo16,Kwiat21}, respectively.
In terms of our notations, in the experiment of ref \cite{Guo16},
the single-traversal photon loss is $\gamma=0.4$,
and the single-pass postselection probability is $p_{d1}=0.1$.
As a result, the signal amplification is $\eta=2.40\pm 0.12$,
and the SNR can indeed surpass that of the single-pass
weak value measurement and the conventional measurement.
In the experiment of ref \cite{Kwiat21}, 
which demonstrated the pulse-recycling scheme,
the photon loss fraction per traversal is $\gamma=0.16$,
and the single-pass postselection probability is $p_{d1}=0.03$.
Under this improved condition, the achieved signal amplification is $\eta=4.4\pm 0.2$,
and the SNR is enhanced by an improvement factor of $\widetilde{R}_{S/N}=2.10\pm 0.06$,
compared with both the single-pass weak-value measurement and the conventional measurement.

In our case, for small $\theta$, we have $\tilde{\theta}=\theta/\sqrt{4p_{d1}}$.
If we assume $p_{d1}=0.03$, which corresponds to $\theta=10^{-3}$ and $\beta=0.11\pi$,
we have $\eta=\tilde{\theta}/\theta=2.89$.
For the SNR, following ref \cite{Kwiat21}, we assume $\gamma=0.16$.
In order to satisfy the impedance matching condition,
we set $r=0.9$ and $\varphi_0=1.945\pi$ since $\chi=0.055\pi$.
Under these choices, we have $\widetilde{R}_{S/N}=2.24$.
Therefore, for the optical sensing of magnetic field under present study,
the signal amplification and SNR enhancement
are quantitatively consistent with those of the power-recycled
interferometric weak-value-based beam-deflection measurement \cite{Guo16,Kwiat21}.
In order to achieve higher SNR, the key requirement is to reduce photon loss.
Only under this condition, one can decrease the postselection probability $p_{d1}$,
by modulating $\beta$, to realize signal amplification and enhance SNR, simultaneously.

To summarize, based on the MZI, we have proposed and analyzed the technique
of postselected amplification combined with continuous-wave power recycling,
for enhanced-sensitivity optical sensing of weak magnetic fields.
The interferometric setup allows not only path-information selection for amplification of signal,
but also coherent multiple interactions of the recycled light with the optical magnetometer.
As a result, while each postselected photon holds an amplified signal parameter,
the total postselected output light intensity
is much stronger than the single-pass postselected light.
Combination of these two advantages can result in an enhanced SNR.
In theory, both the signal amplification and the SNR can attain large values.
Nevertheless, special care must be taken owing to practical limitations,
such as photon loss in the recycling loop.
The proposed scheme shows promising prospects for applications
in precision measurement of weak magnetic fields.

\vspace{0.5cm}


\end{document}